\documentclass[sigconf,screen]{acmart}
\usepackage{multicol,multirow}
\AtBeginDocument{%
  }

\begin{document}

\title{Evaluating the Efficacy of Large Language Models for Generating Fine-Grained Visual Privacy Policies in Homes}

\author{Shuning Zhang}
\orcid{0000-0002-4145-117X}
\email{zsn23@mails.tsinghua.edu.cn}
\affiliation{%
  \institution{Tsinghua University}
  \city{Beijing}
  \country{China}
}

\author{Ying Ma}
\orcid{0000-0001-5413-0132}
\affiliation{%
  \department{School of Computing and \\Information Systems}
  \institution{University of Melbourne}
  \city{Melbourne}
  \country{Australia}}
\email{ying.ma1@student.unimelb.edu.au}

\author{Xin Yi}
\orcid{0000-0001-8041-7962}
\authornote{Corresponding author.}
\email{yixin@tsinghua.edu.cn}
\affiliation{
    \institution{Tsinghua University}
    \city{Beijing}
    \country{China}
}

\author{Hewu Li}
\orcid{0000-0002-6331-6542}
\email{lihewu@cernet.edu.cn}
\affiliation{
    \institution{Tsinghua University}
    \city{Beijing}
    \country{China}
}

\renewcommand{\shortauthors}{Zhang et al.}

\begin{abstract}
  The proliferation of visual sensors in smart home environments, particularly through wearable devices like smart glasses, introduces profound privacy challenges. Existing privacy controls are often static and coarse-grained, failing to accommodate the dynamic and socially nuanced nature of home environments. This paper investigates the viability of using Large Language Models (LLMs) as the core of a dynamic and adaptive privacy policy engine. We propose a conceptual framework where visual data is classified using a multi-dimensional schema that considers data sensitivity, spatial context, and social presence. An LLM then reasons over this contextual information to enforce fine-grained privacy rules, such as selective object obfuscation, in real-time. Through a comparative evaluation of state-of-the-art Vision Language Models (including GPT-4o and the Qwen-VL series) in simulated home settings , our findings show the feasibility of this approach. The LLM-based engine achieved a top machine-evaluated appropriateness score of 3.99 out of 5, and the policies generated by the models received a top human-evaluated score of 4.00 out of 5.
\end{abstract}


\begin{teaserfigure}
  \centering
  \includegraphics[width=\textwidth]{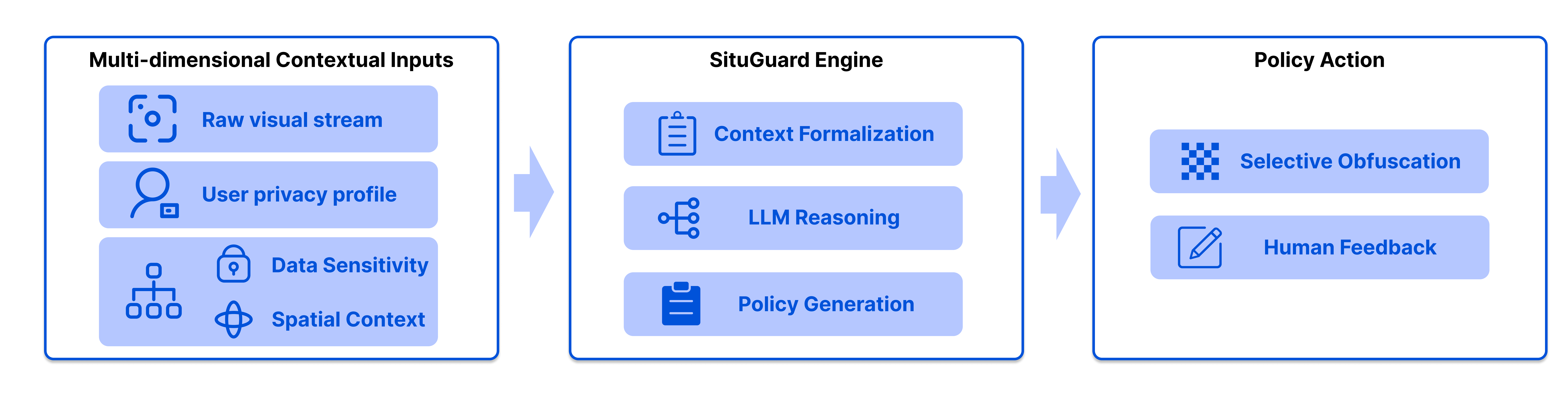}
  \caption{The conceptual framework of SituGuard. The diagram shows the end-to-end process, from ingesting multi-dimensional contextual inputs and user profiles to the LLM-powered engine generating an actionable privacy policy, which is then validated by both machine and human evaluation.}
  \label{fig:teaser}
\end{teaserfigure}

\maketitle

\section{Introduction}

The proliferation of wearable computing, particularly smart glasses equipped with always-on visual sensors, marks a paradigm shift in personal data collection~\cite{abraham2024you}. As these devices become integrated into daily life, they are increasingly utilized within the home, a setting traditionally regarded as a private sanctuary. The continuous and unobtrusive recording capability of smart glasses generates an unprecedented stream of highly personal visual data, creating profound and complex privacy challenges for inhabitants and their guests who struggle to understand and manage how their visual information is being captured and used~\cite{zhang2025through,gallardo2023speculative}.

The central challenge lies in the fundamental mismatch between the static nature of existing privacy controls and the dynamic, socially-nuanced context of home life. An individual's privacy preferences are not absolute~\cite{asthana2024know,zhang2024ghost}. They are contingent upon a fluid interplay of factors, including their location within the home~\cite{song2020m}, the activity being performed~\cite{barbosa2019if}, and the social context~\cite{alom2019helping}. Current control mechanisms, typically limited to coarse-grained, binary permissions, are ill-equipped to manage this complexity. This inadequacy creates a critical dilemma: users desire fine-grained, context-sensitive control over their data, but the cognitive burden of manually configuring policies for every possible situation is prohibitive.

Previous attempts to address this issue have revealed a significant research gap~\cite{venugopalan2024aragorn,abraham2024you,felt2012android,felt2012ask}. Early access control models, adapted from mobile computing, fail to account for the continuous and passive nature of wearable sensors~\cite{felt2012android,felt2012ask}. While more recent context-aware systems have emerged~\cite{abraham2024you,venugopalan2024aragorn}, they often lack true object-level granularity or still demand significant manual intervention from the user to define privacy policies. This leaves a critical need for an intelligent system capable of automatically inferring situational context and dynamically enforcing fine-grained privacy rules without overwhelming the user. To address this gap, we explore the following central research question: \textbf{To what extent are LLMs capable of generating appropriate, context-aware privacy rules for visual sensors in diverse home settings?}

We propose a conceptual framework where an LLM acts as the central reasoning component. This engine receives contextual inputs classified by a multi-dimensional schema, which includes the sensitivity of visible data (e.g., a person's face vs. a piece of furniture), the spatial context (e.g., kitchen vs. home office) and the social presence (e.g., user is alone vs. with guests). Using this information, the LLM generates and applies fine-grained privacy policies in real-time, such as selectively obfuscating specific objects captures by a smart glass camera.

Our evaluation compares the performance of state-of-the-art LLMs, including GPT-4, Llama 3 and Gemini, on the core task of generating context-appropriate privacy rules for a variety of simulated private data settings within a home environment. We found all models are feasible of generating satisfying data flows, providing empirical evidence of the feasibility of this approach. Our findings provide empirical evidence that this approach is feasible, as all evaluated models were capable of generating satisfactory policies. Specifically, \textit{qwen-2.5-vl-72b} model emerged as the top performance, achieving 3.99 out of 5 in machine evaluations, followed closely by \textit{qwen-vl-max} with a score of 3.96. In contrast, \textit{gpt-4o} received the lowest average score of 3.28. The policies generated by the models were also assessed by human evaluators, who assigned an average appropriateness score of 3.78 out of 5 across all models. To sum up, this paper makes the following contributions:

$\bullet$ proposed a conceptual framework that uses an LLM to automatically generate visual privacy rules based on users' preferences.

$\bullet$ provided a comparative evaluation of mainstream LLMs for the task of contextual privacy rule generation in home environments.

$\bullet$ presented empirical evidence of LLMs' feasibility, validating it as a viable tool for intelligent privacy control.

\section{Related Work}

This section provides a comprehensive review of existing literature by first examining privacy management in ubiquitous computing on home environments. Second, we discuss context-aware computing to modeling and utilizing context for system adaptation. Third, we will analyze access control models and visual privacy-preserving techniques.

\subsection{Privacy Management in Home Environments}
Managing privacy in ubiquitous computing (UbiComp) settings such as smart home environments presents unique challenges due to the inherently contextual, dynamic, and personal nature of these spaces. \citet{liu2022development} articulated that privacy protection in smart homes must address the entire data lifecycle, encompassing collection, use, storage, and sharing, with particular emphasis on the context-dependency of privacy needs and user identification protocols. This aligns with~\citet{chhetri2022mute} findings that various privacy concerns can be classified based on established taxonomies, highlighting categories such as information collection, information processing, and information dissemination. These concerns are compounded by the pervasive nature of IoT devices, which continuously monitor and interact with inhabitants, emphasizing the need for consumers to consider their personal privacy preferences and actively engage in protective measures~\cite{zheng2018user}.

\subsection{Access Control of Privacy-preserving Techniques}

Access control systems for smart glasses evolved from paradigms for mobile devices, which initially relied on install-time and later run-time permission models~\cite{felt2012ask}. Install-time permissions require user consent at the moment of application installation, which was criticized for its lack of transparency~\cite{felt2012android}. Run-time permissions reduced the potential for over-privileged applications through granular or one-time access to a camera or location data~\cite{wijesekera2017feasibility}. However, run-time model remains insufficient for the unique continuous, passive and often surreptitious data collection on smart glasses~\cite{wijesekera2018contextualizing}. In response to these limitations, dynamic permission systems occurred, highlighting adapting control to contextual cues~\cite{baarslag2016negotiation,wijesekera2018contextualizing}. These systems infer user intent and reduce permission requests through machine learning, improving user engagement and protection~\cite{zhou2016enhancing}. Systems like PriView further enhance this by providing privacy visualization as feedback~\cite{prange2021priview}. However, while effective at a holistic level, many context-aware models under-explore the fine-grained, object-level control for many real-world scenarios~\cite{abraham2024you,zhou2016enhancing}.


This necessity for greater precision inspired fine-grained permission, which contained (1) \textit{modifying the binary choices to allow for more control}~\cite{abraham2024you,nair2023going} and (2) \textit{enabling users to manually design the object to permit}~\cite{venugopalan2024aragorn,raval2016you,lee2024priviaware,jana2013enabling,hasan2020automatically,shu2018cardea,aditya2016pic}. Modification of binary choices started from location data, where iOS and Android allowed users to select between approximate or precise location information when using applications~\cite{google_android_2022,fu2014general,apple_ios_2022,ma2024understanding}. Incognito~\cite{nair2023going} used differential privacy for protecting users' personal privacy in VR, with slider to balance the parameters between privacy and utility. Abraham et al.~\cite{abraham2024you} designed a slider for fine-grained permission control for camera and microphones, allowing users to select discrete balancing levels. However how to enable finer control beyond slider-based levels and how can users understand and learn worth exploration.

Other researchers started from automatically recognizing bystanders to filter out these people~\cite{hasan2020automatically}, and then transferred to context-aware recognition of objects~\cite{shu2018cardea,aditya2016pic}, enabling automatic control on photos. However, these methods only considered a single class or few salient classes one at a time, omitting that many tasks in daily home activities needed more than one saliency object to function. Another line of researchers enabled users to manually configure the policy~\cite{lee2024priviaware} or set the saliency object to block out~\cite{raval2016you} or retain~\cite{jana2013enabling}. However, manually setting the blocking object~\cite{raval2016you} or configuring privacy~\cite{lee2024priviaware} is costly for users, especially for privacy protection, which is often the side task beyond primary goals. Aragorn built on previous work by supporting automatically recognizing objects and enabling users to select which object to retain~\cite{venugopalan2024aragorn}, leaving the space to explore how to efficiently configuring multiple objects.

\section{Conceptual Framework for LLM-based Policy Generation}

To address the challenge of dynamic privacy management, we propose a conceptual framework centered on an LLM-powered policy engine (Figure~\ref{fig:teaser}). This architecture is designed to translate complex, multi-dimensional environmental context into specific, actionable privacy rules. The framework consists of two interconnected components: a multi-dimensional privacy schema that structures contextual information and an LLM-powered engine that reasons over this information.

\textbf{Multi-Dimensional Privacy Preference}
The foundation of our framework is a comprehensive schema that classifies situational data along several orthogonal dimensions, moving beyond simple object detection to capture the rich context of the home environment. This schema is critical for providing the LLM with the structured input necessary for nuanced reasoning. The primary dimensions include:

$\bullet$ Data Sensitivity: This dimension categorizes detected objects or events on a spectrum of privacy sensitivity. This ranges from low sensitivity, such as inanimate household objects, to high sensitivity, which includes personally identifiable information (PII), health-related inferences, financial documents, and detailed behavioral logs in privacy zones. We classified data to three different sensitivity levels: low sensitivity, middle sensitivity and high sensitivity according to prior work~\cite{orekondy2017towards,wang2023modeling}. 

$\bullet$ Spatial Context: Drawing upon established principles in Ubiquitous Computing (Ubicomp) literature, this dimension defines the physical location where data is being captured. The home is partitioned into distinct spatial zones, such as a sleeping space, working space or living space, each associated with different baseline privacy expectations. We classified the space into sleeping space, working space and living space.

$\bullet$ Contextual Modifiers: This dimension accounts for transient social and temporal factors that dynamically alter privacy requirements. Key modifiers include social presence (e.g., residents only, presence of guests, presence of children) and temporal factors (e.g., time of day, day of week), which can significantly influence the appropriateness of data collection.

\textbf{LLM-Powered Adaptive Engine} The policy engine operationalizes the schema by using an LLM to perform real-time privacy policy generation. The process follows a structured pipeline:

$\bullet$ Context Formalization: Raw data from environmental sensors (e.g., a list of detected objects from a visual stream, location data from an indoor positioning system) is first parsed. This information is then structured according to the multi-dimensional schema, creating a formalized representation of the current situation.

$\bullet$ Prompt Engineering: We assembled a carefully constructed prompt and sent it to the LLM. The prompt includes: (1) the formalized situational context from the previous step, including the explicit privacy rules previously defined by the user, and a directive for the LLM to perform zero-shot reasoning and generate data acceptability rules for the privacy policy.

$\bullet$ Policy Generation and Action: The LLM processes the prompt and generates a policy output, in a structured JSON format. For each sensor type, the LLM generates the acceptability separately. The system would latter use the generated policy for actionable sensor control, and incorporate user feedback to continuously refine its rule set.

\section{Evaluation Methodology}

To empirically validate the capability of LLMs for this task, we designed a comprehensive evaluation methodology focused on assessing policy generation accuracy and user-perceived appropriateness.

\begin{figure}
    \centering
    \includegraphics[width=0.5\textwidth]{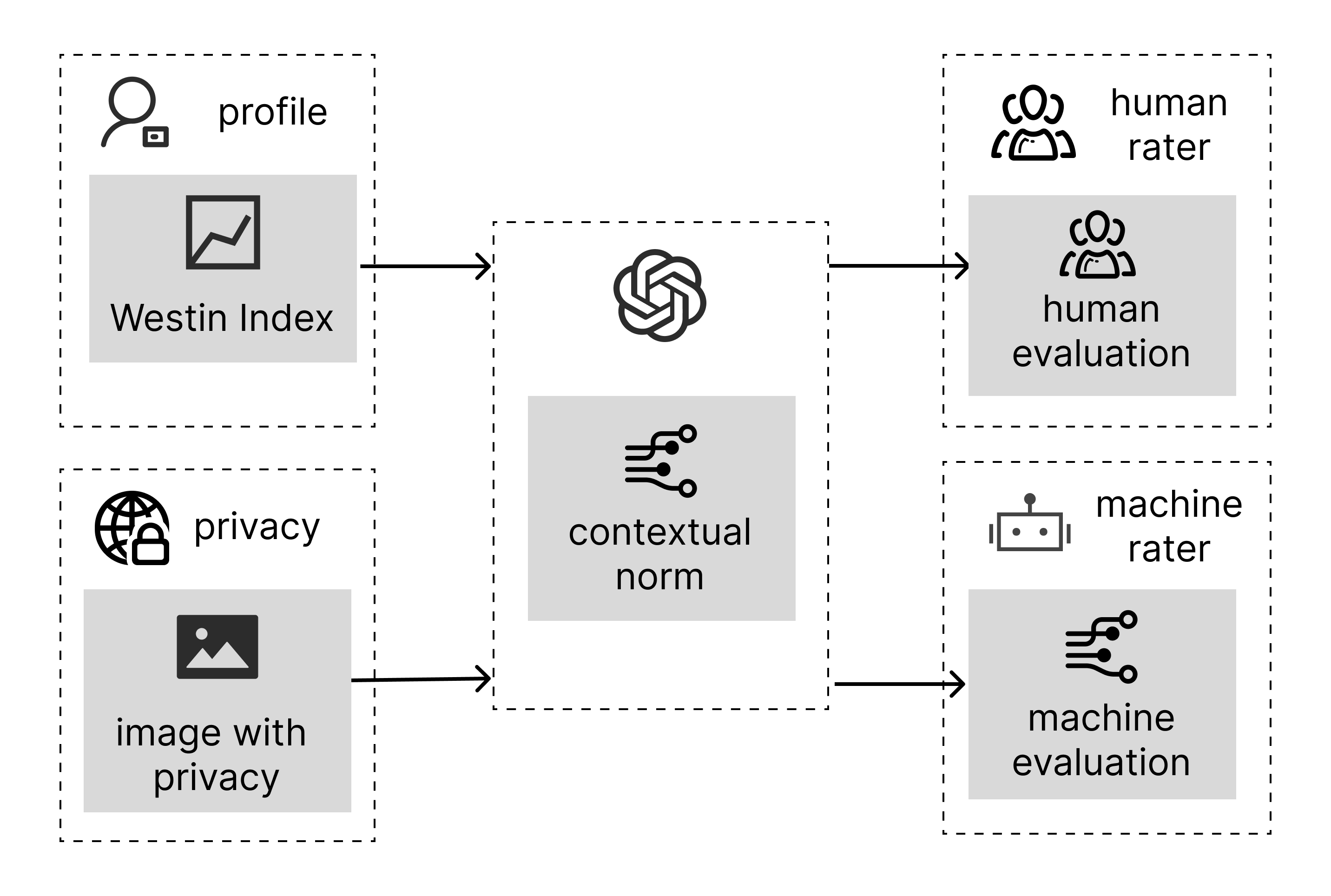}
    \caption{Evaluation protocol for Situguard.}
    \label{fig:situguard}
\end{figure}

\subsection{Models Compared}

Our comparative analysis evaluated five prominent Vision Language Models (VLMs), selected to represent the current state-of-the-art and to provide a spectrum of model scales and architectures. The selection includes models from different leading developers, enabling a robust assessment of their capabilities in the specialized domain of dynamic privacy policy generation.

$\bullet$ GPT-4o (OpenAI): It is recognized for its advanced reasoning, instruction-following capabilities, and its native integration of text, audio and vision processing.

$\bullet$ Qwen-VL series: We included a comprehensive range of models from the Qwen-VL series: \textit{qwen-vl-max}, a high-performance large vision-language model representing the Qwen series' capabilities, qwen2.5-vl-72b, qwen2.5-vl-32b, and qwen2.5-vl-7b, which represent different scale, from a 7b parameter model to a 72b parameter model.

\subsection{Evaluation Dataset}

We selected three datasets, with two featuring general (and also containing home environment) images with privacy annotations~\cite{xu2023dipa,xu2024dipa2}, and the third featuring privacy annotations on in-home videos~\cite{wu2020privacy}.

$\bullet$ DIPA~\cite{xu2023dipa}: it collected 1,495 images with 5,671 annotations from LVIS~\cite{gupta2019lvis} and OpenImages~\cite{kuznetsova2020open} dataset, featuring 22 primary privacy categories and other categories annotated by both Japan and UK annotators.

$\bullet$ DIPA2~\cite{xu2024dipa2}: it collected 1,304 images with 3,347 objects and 5,897 annotations, from both Japan and UK annotators. 

$\bullet$ PA-HMDB51~\cite{wu2020privacy}: it has 5 attributes skin color, gender, face, nudity and personal relationship on 515 videos selected from HMDB51.

\subsection{Task Design}

Our task design simulates a realistic, multi-stage process for dynamic privacy control, mirroring the conceptual framework illustrated in Figure~\ref{fig:situguard}. The methodology is structured to first generate a context-rich scenario and then evaluate a VLM's ability to generate an appropriate, user-aligned privacy policy for that scenario. The process unfolds as follows:

$\bullet$ Scenario and profile generation: Each evaluation task begins with the generation of a unique scenario. A visual input, such as a video frame from a dataset, is selected. This visual scene is then paired with a randomly assigned user privacy profile. These profiles are derived from established frameworks like Westin's Privacy Index~\cite{kumaraguru2005privacy}, categorizing users into archetypes such as \textit{``privacy fundamentalist''}, \textit{``privacy pragmatist''}, or \textit{``privacy unconcerned''} to dictate their sensitivity to data collection. We selected the representative scenarios from the prior work~\cite{arakawa2024prism,arakawa2023prism,mahmud2024actsonic,steil2019privaceye}, featuring four types of tasks: daily health monitoring and behavior management~\cite{arakawa2024prism,arakawa2023prism,mahmud2024actsonic,steil2019privaceye}, household and lifestyle management~\cite{arakawa2024prism,arakawa2023prism}, social interaction and contextual awareness~\cite{mahmud2024actsonic,steil2019privaceye}, multimodal learning and work assistance~\cite{mahmud2024actsonic,steil2019privaceye}.

$\bullet$ VLM policy proposal: The core of the task involves the VLM under evaluation. The model is presented with the visual input and the corresponding user profile. It is then prompted to perform two key actions: first, to identify and recognize objects, individuals, and other potentially sensitive elements within the scene; and second, to reason over this visual analysis in conjunction with the user's privacy profile. Based on this reasoning, the VLM generates a fine-grained privacy policy proposal in a structured JSON format. This proposal explicitly recommends which elements within the scene (e.g., faces, documents, specific objects) should be anonymized or obfuscated.

$\bullet$ Appropriateness evaluation: The final stage is the evaluation of the VLM-generated policy. The appropriateness of the model's decision-making is assessed using a two-aspect approach. First, the generated policy is evaluated using LLMs to calculate an appropriateness score. Second, human evaluators are recruited to provide a qualitative user satisfaction rating. The LLMs and the participants are shown the scenario and the model's proposed anonymization action, then rate its appropriateness on a Likert scale (1 to 5), where 1 indicated very inappropriate, and and 5 indicated very appropriate, providing a crucial measure of how well the model's logic aligns with human expectations. We recruited 36 human evaluators for performing the evaluation on each evaluated dataset. 





\section{Results}

Our empirical evaluation yielded clear insights into the capabilities and comparative performance of contemporary LLMs for dynamic privacy policy generation. 

\subsection{Policy Generation Appropriateness}

The quantitative accuracy of our evaluation are presented in Table~\ref{tab:eval_comparison}. The dataset reveals a clear hierarchy of performance, with \textit{qwen2.5-vl-72b} model emerging as the top-performing model, achieving the highest average appropriateness score of 3.99. It was closely followed by \textit{qwen-vl-max}, which obtained an average score of 3.96. These results indicate that models with larger parameter counts and specialized vision-language architectures generate policies that are more contextually suitable and comprehensive. The outputs from these two models consistently showed a grasp of the requisite components of a standard privacy policy, such as the explicit statement of data collection purposes, usage modalities and user rights.

In contrast, \textit{gpt-4o} got the lowest average score of 3.28. While its generated policies were generally relevant, they were observed to be less consistent in meeting the high standard of appropriateness required for this task. The models of intermediate scale, \textit{qwen2.5-vl-7b} and \textit{qwen2.5-vl-32b} achieved average scores of 3.51 and 3.69 respectively, suggesting a positive correlation of model size and the ability to handle the nuances of generating appropriate legal and informational documents.

Furthermore, we observed a notable variation in performance in datasets. All models, and the \textit{qwen} series in particular, exhibited significantly higher scores on the PA-HMDB51 dataset. On this dataset \textit{qwen2.5-vl-72b} attained a peak score of 4.33, and even \textit{qwen2.5-vl-7b}'s score rose to 3.95. This suggests that the nature and complexity of the scenarios within PA-HMDB51 align better with the model's capabilities, allowing them to show a greater proficiency. 


\begin{table*}[!htbp]
\centering
\caption{Policy generation appropriateness scores (out of 5) with machine and human evaluation.}
\label{tab:eval_comparison}
\setlength{\tabcolsep}{6pt} 
\begin{tabular}{l cc cc cc cc}
\toprule
\multirow{2}{*}{\textbf{Model}} & \multicolumn{2}{c}{\textbf{DIPA}} & \multicolumn{2}{c}{\textbf{DIPA2}} & \multicolumn{2}{c}{\textbf{PA-HMDB51}} & \multicolumn{2}{c}{\textbf{Avg. Score}} \\
\cmidrule(lr){2-3} \cmidrule(lr){4-5} \cmidrule(lr){6-7} \cmidrule(lr){8-9}
& Mach. & Human & Mach. & Human & Mach. & Human & Mach. & Human \\
\midrule
gpt-4o         & 3.22 & 3.14 & 3.22 & 3.24 & 3.39 & 3.36 & 3.28 & 3.25 \\
qwen-vl-max    & 3.75 & 3.54 & 3.82 & 3.75 & 4.31 & 4.15 & 3.96 & 3.81 \\
qwen2.5-vl-7b  & 3.27 & 3.40 & 3.30 & 3.55 & 3.95 & 3.87 & 3.51 & 3.61 \\
qwen2.5-vl-32b & 3.47 & 3.45 & 3.59 & 3.66 & 4.00 & 3.92 & 3.69 & 3.68 \\
qwen2.5-vl-72b & 3.77 & 3.85 & 3.87 & 3.91 & 4.33 & 4.24 & 3.99 & 4.00 \\
\bottomrule
\end{tabular}
\end{table*}

\subsection{Case Study}

To instantiate the quantitative findings, we present a representative case study from the PA-HMDB51 dataset. This case examines the performance of the qwen2.5-vl-7b model when presented with a video frame depicting a child performing a cartwheel, with an adult partially visible in the background. The task was contextualized by the \textit{``Privacy Fundamentalist''} user profile, which mandates a high level of sensitivity towards any data that could reveal identity, location, or personal activities.

In its initial proposal stage, the model correctly identified the primary subjects of privacy concern. It recommended anonymizing two distinct regions:

$\bullet$ \textit{The child}: The model reasoned that \textit{``The child’s face could potentially reveal identifiable features that might compromise their privacy.''}

$\bullet$ \textit{The adult figure}: For this subject, the model justified anonymization by noting, \textit{``This adult figure may contain details such as facial recognition patterns due to high sensitivity towards identity-related data.''}

This proposal was subsequently evaluated and awarded an appropriateness score of 4 out of 5. The justification for this score highlighted that the model's recommendations were highly aligned with the stringent requirements of the user profile. The model successfully prioritized the anonymization of human subjects, which are the most critical elements for a user concerned with identity privacy. The reasoning provided was directly relevant to the profile's stated sensitivities regarding facial features and identity. The evaluation noted a minor limitation: the proposal could have achieved perfect completeness by also considering the anonymization of other potentially identifying environmental markers, such as logos or distinctive text, had they been present. 

\subsection{Ablation Study}

We conducted an ablation study to validate the contribution of each key component in our framework. By systematically removing or simplifying modules, we demonstrate that our proposed multi-dimensional context and user profile integration are essential for generating high-quality privacy policies. We compare our full framework SituGuard against three ablated versions: no-context model, which removes the multi-dimensional privacy schema; simplified-context model, which provides only the data sensitivity dimension of the context; and profile-agnostic model, which remove the user's privacy profile (Table~\ref{tab:ablation}). 

\textbf{Results of Ablation Study} The results of our ablation study provide compelling evidence for the integral function of each component in our framework. The SituGuard model significantly outperforms all ablated versions. The most substantial performance degradation is observed in the no-context model, which demonstrates that without a structured, multi-dimensional understanding of the environment, the LLM is incapable of reasoning effectively about nuanced privacy requirements. The simplified-context model shows a moderate decline, indicating that while data sensitivity is important, policies are less appropriate without spatial and social cues. Finally, the profile-agnostic model performs reasonably well, but its lower scores confirm that tailoring policies to an individual's declared privacy preferences is critical for achieving the highest levels of appropriateness and user satisfaction (see Table~\ref{tab:ablation}). 

\begin{table}[!htbp]
\centering
\caption{Ablation study results on the PA-HMDB51 dataset generated by qwen-vl-max.}
\label{tab:ablation}
\begin{tabular}{llcc}
\toprule
\textbf{Model Variant} & \textbf{Machine Eval.} & \textbf{Human Eval.} \\
\midrule
SituGuard & 4.31 & 4.15 \\
No-Context Model & 2.89 & 2.75 \\
Simplified-Context Model & 3.61 & 3.48 \\
Profile-Agnostic Model & 3.82 & 3.71 \\
\bottomrule
\end{tabular}
\end{table}

\section{Discussion and Future Work}

\subsection{Cultural Nuances of Contextual Policy}

Privacy is inherently sensitive to cultures and social norms~\cite{xu2023dipa,xu2024dipa2}, where people with different cultures and norms may have different preferences, which LLMs may could not perfectly handle. However, our method models the human's preference as a priori condition during the input of the adaptive policy generation, which could potentially adapt to different preferences than static control~\cite{venugopalan2024aragorn}. Besides, different cultures may have different preferences of control, agency, and automation. While our current system uses AI for delegating the task~\cite{zhang2019proactive}, the system could also explore different granularity and role assignment regarding the control, especially for different cultures~\cite{zhang2024adanonymizer}. 

\subsection{Feasibility and Implications of SituGuard}

SituGuard uses a localized vision-based model for contextually setting the privacy policy, which potentially mitigate the risks of passing the private information, especially the visual information online~\cite{zhao2025visual}. Our experiments verified that even a 7b model could achieve comparable capabilities for contextually anonymizing the information, which is potentially feasible for localized processing. However, the processing requirements and hardware requirements for localized processing may need further exploration and engineering.

Besides privacy control, it is also important to adopt privacy visualization~\cite{lee2024priviaware,prange2021priview} as a manner for meaningful explanation. Users may need the explanation about the data control, especially towards which data SituGuard proactively anonymizes and which it did not. This could be achieved through on-device overlay~\cite{prange2021priview}, post-usage report~\cite{lee2024priviaware}. SituGuard's visualization should also be succinct, human-intelligible justifications rather than lengthy privacy policies~\cite{zhang2025privcaptcha} for users' engagement and sense-making. Some users may desire manual control, which granted them with higher agency~\cite{zhang2019proactive}, in comparison to those prioritizing automatic control. Therefore, SituGuard could be integrated with user-initiated permission control systems~\cite{abraham2024you,venugopalan2024aragorn} to present users with different control settings.

SituGuard is designed to function in home environment, which has rich private information, and has complicated privacy risks~\cite{orekondy2017towards,wu2020privacy}. It could be deployed as a OS-level application of smart glasses, released by the service provider to protect users' privacy from being invaded by third-party applications. Besides, its underlying policy engine could also inspire the privacy control of inter-personal privacy risks~\cite{zhou2024bring}, privacy risks in office environments~\cite{li2023occupant} and other risks regarding social media environments~\cite{tomekcceprivate,ma2025raising}. Its method of aligning with users' privacy preferences through prompts also coincided with prior work~\cite{zhang2024privacy}, which proves a promising path for easing users' control efforts and achieving automatic privacy redaction~\cite{zhang2019proactive}. 

Smart home could have multiple stakeholders, including guests and hosts~\cite{geeng2019s}, workers~\cite{he2025exploring} and different home members. Adapting to different requirements of different home members not only need SituGuard to respect the wearer's privacy preference, but also respect the bystanders' privacy preference~\cite{ahmad2020tangible,o2023privacy}, as they may have different perceptions towards recording, among which even some may be conflicting~\cite{alshehri2023exploring}, especially in smart home context. 



\subsection{Limitations and Future Work}

We acknowledge despite the promising direction we showed, our work has several limitations. First, our evaluation was conducted using curated datasets within a controlled environment. While this ensures reproducibility and follows prior guidance~\cite{hu2024exploring,zhang2025interveeg}, it may simplifies the unpredictability of real-world domestic life, where evolving routines and complex social negotiations~\cite{zhou2024bring} may complicate the trade-offs. The future work could embed these methods in the real world scenarios and explore users' experience. Besides, we used LLMs for evaluation. Although it is commonly employed in AI literature~\cite{zheng2023judging,chen2024mllm}, it may be different from the realistic human evaluation~\cite{zhang2024ghost}, and we regarded the human-involved evaluation as the future step. We also used a limited set of LLMs for evaluation. Although our models considered different parameter sizes, different brands, both open-sourced and close-sourced models, future work could explore more models and put further efforts into evaluating localized models.

\begin{acks}
This work was supported by the Natural Science Foundation of China under Grant No. 62472243 and 62132010.
\end{acks}

\bibliographystyle{ACM-Reference-Format}
\bibliography{sample-base}


\begin{thebibliography}{60}


\ifx \showCODEN    \undefined \def \showCODEN     #1{\unskip}     \fi
\ifx \showISBNx    \undefined \def \showISBNx     #1{\unskip}     \fi
\ifx \showISBNxiii \undefined \def \showISBNxiii  #1{\unskip}     \fi
\ifx \showISSN     \undefined \def \showISSN      #1{\unskip}     \fi
\ifx \showLCCN     \undefined \def \showLCCN      #1{\unskip}     \fi
\ifx \shownote     \undefined \def \shownote      #1{#1}          \fi
\ifx \showarticletitle \undefined \def \showarticletitle #1{#1}   \fi
\ifx \showURL      \undefined \def \showURL       {\relax}        \fi
\providecommand\bibfield[2]{#2}
\providecommand\bibinfo[2]{#2}
\providecommand\natexlab[1]{#1}
\providecommand\showeprint[2][]{arXiv:#2}

\bibitem[Abraham et~al\mbox{.}(2024)]%
        {abraham2024you}
\bibfield{author}{\bibinfo{person}{Melvin Abraham}, \bibinfo{person}{Mark Mcgill}, {and} \bibinfo{person}{Mohamed Khamis}.} \bibinfo{year}{2024}\natexlab{}.
\newblock \showarticletitle{What You Experience is What We Collect: User Experience Based Fine-Grained Permissions for Everyday Augmented Reality}. In \bibinfo{booktitle}{\emph{Proceedings of the CHI Conference on Human Factors in Computing Systems}}. \bibinfo{pages}{1--24}.
\newblock


\bibitem[Aditya et~al\mbox{.}(2016)]%
        {aditya2016pic}
\bibfield{author}{\bibinfo{person}{Paarijaat Aditya}, \bibinfo{person}{Rijurekha Sen}, \bibinfo{person}{Peter Druschel}, \bibinfo{person}{Seong Joon~Oh}, \bibinfo{person}{Rodrigo Benenson}, \bibinfo{person}{Mario Fritz}, \bibinfo{person}{Bernt Schiele}, \bibinfo{person}{Bobby Bhattacharjee}, {and} \bibinfo{person}{Tong~Tong Wu}.} \bibinfo{year}{2016}\natexlab{}.
\newblock \showarticletitle{I-pic: A platform for privacy-compliant image capture}. In \bibinfo{booktitle}{\emph{Proceedings of the 14th annual international conference on mobile systems, applications, and services}}. \bibinfo{pages}{235--248}.
\newblock


\bibitem[Ahmad et~al\mbox{.}(2020)]%
        {ahmad2020tangible}
\bibfield{author}{\bibinfo{person}{Imtiaz Ahmad}, \bibinfo{person}{Rosta Farzan}, \bibinfo{person}{Apu Kapadia}, {and} \bibinfo{person}{Adam~J Lee}.} \bibinfo{year}{2020}\natexlab{}.
\newblock \showarticletitle{Tangible privacy: Towards user-centric sensor designs for bystander privacy}.
\newblock \bibinfo{journal}{\emph{Proceedings of the ACM on Human-Computer Interaction}} \bibinfo{volume}{4}, \bibinfo{number}{CSCW2} (\bibinfo{year}{2020}), \bibinfo{pages}{1--28}.
\newblock


\bibitem[Alom et~al\mbox{.}(2019)]%
        {alom2019helping}
\bibfield{author}{\bibinfo{person}{Md~Zulfikar Alom}, \bibinfo{person}{Barbara Carminati}, {and} \bibinfo{person}{Elena Ferrari}.} \bibinfo{year}{2019}\natexlab{}.
\newblock \showarticletitle{Helping users managing context-based privacy preferences}. In \bibinfo{booktitle}{\emph{2019 IEEE International Conference on Services Computing (SCC)}}. IEEE, \bibinfo{pages}{100--107}.
\newblock


\bibitem[Alshehri et~al\mbox{.}(2023)]%
        {alshehri2023exploring}
\bibfield{author}{\bibinfo{person}{Ahmed Alshehri}, \bibinfo{person}{Eugin Pahk}, \bibinfo{person}{Joseph Spielman}, \bibinfo{person}{Jacob~T Parker}, \bibinfo{person}{Benjamin Gilbert}, {and} \bibinfo{person}{Chuan Yue}.} \bibinfo{year}{2023}\natexlab{}.
\newblock \showarticletitle{Exploring the negotiation behaviors of owners and bystanders over data practices of smart home devices}. In \bibinfo{booktitle}{\emph{Proceedings of the 2023 CHI Conference on Human Factors in Computing Systems}}. \bibinfo{pages}{1--27}.
\newblock


\bibitem[Android(2022)]%
        {google_android_2022}
\bibfield{author}{\bibinfo{person}{Google Android}.} \bibinfo{year}{2022}\natexlab{}.
\newblock \bibinfo{title}{Request location permissions}.
\newblock
\urldef\tempurl%
\url{https://developer.android.com/training/location/permission}
\showURL{%
\tempurl}
\newblock
\shownote{Accessed: 2025-01-14}.


\bibitem[Arakawa et~al\mbox{.}(2024)]%
        {arakawa2024prism}
\bibfield{author}{\bibinfo{person}{Riku Arakawa}, \bibinfo{person}{Hiromu Yakura}, {and} \bibinfo{person}{Mayank Goel}.} \bibinfo{year}{2024}\natexlab{}.
\newblock \showarticletitle{PrISM-Observer: Intervention agent to help users perform everyday procedures sensed using a smartwatch}. In \bibinfo{booktitle}{\emph{Proceedings of the 37th Annual ACM Symposium on User Interface Software and Technology}}. \bibinfo{pages}{1--16}.
\newblock


\bibitem[Arakawa et~al\mbox{.}(2023)]%
        {arakawa2023prism}
\bibfield{author}{\bibinfo{person}{Riku Arakawa}, \bibinfo{person}{Hiromu Yakura}, \bibinfo{person}{Vimal Mollyn}, \bibinfo{person}{Suzanne Nie}, \bibinfo{person}{Emma Russell}, \bibinfo{person}{Dustin~P DeMeo}, \bibinfo{person}{Haarika~A Reddy}, \bibinfo{person}{Alexander~K Maytin}, \bibinfo{person}{Bryan~T Carroll}, \bibinfo{person}{Jill~Fain Lehman}, {et~al\mbox{.}}} \bibinfo{year}{2023}\natexlab{}.
\newblock \showarticletitle{Prism-tracker: A framework for multimodal procedure tracking using wearable sensors and state transition information with user-driven handling of errors and uncertainty}.
\newblock \bibinfo{journal}{\emph{Proceedings of the ACM on Interactive, Mobile, Wearable and Ubiquitous Technologies}} \bibinfo{volume}{6}, \bibinfo{number}{4} (\bibinfo{year}{2023}), \bibinfo{pages}{1--27}.
\newblock


\bibitem[Asthana et~al\mbox{.}(2024)]%
        {asthana2024know}
\bibfield{author}{\bibinfo{person}{Sumit Asthana}, \bibinfo{person}{Jane Im}, \bibinfo{person}{Zhe Chen}, {and} \bibinfo{person}{Nikola Banovic}.} \bibinfo{year}{2024}\natexlab{}.
\newblock \showarticletitle{" I know even if you don't tell me": Understanding Users' Privacy Preferences Regarding AI-based Inferences of Sensitive Information for Personalization}. In \bibinfo{booktitle}{\emph{Proceedings of the 2024 CHI Conference on Human Factors in Computing Systems}}. \bibinfo{pages}{1--21}.
\newblock


\bibitem[Baarslag et~al\mbox{.}(2016)]%
        {baarslag2016negotiation}
\bibfield{author}{\bibinfo{person}{Tim Baarslag}, \bibinfo{person}{Alper~T Alan}, \bibinfo{person}{Richard~C Gomer}, \bibinfo{person}{Ilaria Liccardi}, \bibinfo{person}{Helia Marreiros}, \bibinfo{person}{Enrico~H Gerding}, {and} \bibinfo{person}{MC Schraefel}.} \bibinfo{year}{2016}\natexlab{}.
\newblock \showarticletitle{Negotiation as an interaction mechanism for deciding app permissions}. In \bibinfo{booktitle}{\emph{Proceedings of the 2016 CHI conference extended abstracts on human factors in computing systems}}. \bibinfo{pages}{2012--2019}.
\newblock


\bibitem[Barbosa et~al\mbox{.}(2019)]%
        {barbosa2019if}
\bibfield{author}{\bibinfo{person}{Nat{\~a}~M Barbosa}, \bibinfo{person}{Joon~S Park}, \bibinfo{person}{Yaxing Yao}, {and} \bibinfo{person}{Yang Wang}.} \bibinfo{year}{2019}\natexlab{}.
\newblock \showarticletitle{“what if?” predicting individual users’ smart home privacy preferences and their changes}.
\newblock \bibinfo{journal}{\emph{Proceedings on Privacy Enhancing Technologies}} (\bibinfo{year}{2019}).
\newblock


\bibitem[Chen et~al\mbox{.}(2024)]%
        {chen2024mllm}
\bibfield{author}{\bibinfo{person}{Dongping Chen}, \bibinfo{person}{Ruoxi Chen}, \bibinfo{person}{Shilin Zhang}, \bibinfo{person}{Yaochen Wang}, \bibinfo{person}{Yinuo Liu}, \bibinfo{person}{Huichi Zhou}, \bibinfo{person}{Qihui Zhang}, \bibinfo{person}{Yao Wan}, \bibinfo{person}{Pan Zhou}, {and} \bibinfo{person}{Lichao Sun}.} \bibinfo{year}{2024}\natexlab{}.
\newblock \showarticletitle{Mllm-as-a-judge: Assessing multimodal llm-as-a-judge with vision-language benchmark}. In \bibinfo{booktitle}{\emph{Forty-first International Conference on Machine Learning}}.
\newblock


\bibitem[Chhetri and Motti(2022)]%
        {chhetri2022mute}
\bibfield{author}{\bibinfo{person}{Chola Chhetri} {and} \bibinfo{person}{Vivian Motti}.} \bibinfo{year}{2022}\natexlab{}.
\newblock \showarticletitle{“I mute my echo when I talk politics”: Connecting Smart Home Device Users’ Concerns to Privacy Harms Taxonomy}. In \bibinfo{booktitle}{\emph{Proceedings of the Human Factors and Ergonomics Society Annual Meeting}}, Vol.~\bibinfo{volume}{66}. SAGE Publications Sage CA: Los Angeles, CA, \bibinfo{pages}{2083--2087}.
\newblock


\bibitem[Felt et~al\mbox{.}(2012a)]%
        {felt2012ask}
\bibfield{author}{\bibinfo{person}{Adrienne~Porter Felt}, \bibinfo{person}{Serge Egelman}, \bibinfo{person}{Matthew Finifter}, \bibinfo{person}{Devdatta Akhawe}, {and} \bibinfo{person}{David Wagner}.} \bibinfo{year}{2012}\natexlab{a}.
\newblock \showarticletitle{How to ask for permission}.
\newblock \bibinfo{journal}{\emph{HotSec}} (\bibinfo{year}{2012}), \bibinfo{pages}{0}.
\newblock


\bibitem[Felt et~al\mbox{.}(2012b)]%
        {felt2012android}
\bibfield{author}{\bibinfo{person}{Adrienne~Porter Felt}, \bibinfo{person}{Elizabeth Ha}, \bibinfo{person}{Serge Egelman}, \bibinfo{person}{Ariel Haney}, \bibinfo{person}{Erika Chin}, {and} \bibinfo{person}{David Wagner}.} \bibinfo{year}{2012}\natexlab{b}.
\newblock \showarticletitle{Android permissions: User attention, comprehension, and behavior}. In \bibinfo{booktitle}{\emph{Proceedings of the eighth symposium on usable privacy and security}}. \bibinfo{pages}{1--14}.
\newblock


\bibitem[Fu and Lindqvist(2014)]%
        {fu2014general}
\bibfield{author}{\bibinfo{person}{Huiqing Fu} {and} \bibinfo{person}{Janne Lindqvist}.} \bibinfo{year}{2014}\natexlab{}.
\newblock \showarticletitle{General area or approximate location? How people understand location permissions}. In \bibinfo{booktitle}{\emph{Proceedings of the 13th Workshop on Privacy in the Electronic Society}}. \bibinfo{pages}{117--120}.
\newblock


\bibitem[Gallardo et~al\mbox{.}(2023)]%
        {gallardo2023speculative}
\bibfield{author}{\bibinfo{person}{Andrea Gallardo}, \bibinfo{person}{Chris Choy}, \bibinfo{person}{Jaideep Juneja}, \bibinfo{person}{Efe Bozkir}, \bibinfo{person}{Camille Cobb}, \bibinfo{person}{Lujo Bauer}, {and} \bibinfo{person}{Lorrie Cranor}.} \bibinfo{year}{2023}\natexlab{}.
\newblock \showarticletitle{Speculative privacy concerns about ar glasses data collection}.
\newblock \bibinfo{journal}{\emph{Proceedings on Privacy Enhancing Technologies}} (\bibinfo{year}{2023}).
\newblock


\bibitem[Geeng and Roesner(2019)]%
        {geeng2019s}
\bibfield{author}{\bibinfo{person}{Christine Geeng} {and} \bibinfo{person}{Franziska Roesner}.} \bibinfo{year}{2019}\natexlab{}.
\newblock \showarticletitle{Who's in control? Interactions in multi-user smart homes}. In \bibinfo{booktitle}{\emph{Proceedings of the 2019 CHI conference on human factors in computing systems}}. \bibinfo{pages}{1--13}.
\newblock


\bibitem[Gupta et~al\mbox{.}(2019)]%
        {gupta2019lvis}
\bibfield{author}{\bibinfo{person}{Agrim Gupta}, \bibinfo{person}{Piotr Dollar}, {and} \bibinfo{person}{Ross Girshick}.} \bibinfo{year}{2019}\natexlab{}.
\newblock \showarticletitle{Lvis: A dataset for large vocabulary instance segmentation}. In \bibinfo{booktitle}{\emph{Proceedings of the IEEE/CVF conference on computer vision and pattern recognition}}. \bibinfo{pages}{5356--5364}.
\newblock


\bibitem[Hasan et~al\mbox{.}(2020)]%
        {hasan2020automatically}
\bibfield{author}{\bibinfo{person}{Rakibul Hasan}, \bibinfo{person}{David Crandall}, \bibinfo{person}{Mario Fritz}, {and} \bibinfo{person}{Apu Kapadia}.} \bibinfo{year}{2020}\natexlab{}.
\newblock \showarticletitle{Automatically detecting bystanders in photos to reduce privacy risks}. In \bibinfo{booktitle}{\emph{2020 IEEE Symposium on Security and Privacy (SP)}}. IEEE, \bibinfo{pages}{318--335}.
\newblock


\bibitem[He et~al\mbox{.}(2025)]%
        {he2025exploring}
\bibfield{author}{\bibinfo{person}{Shijing He}, \bibinfo{person}{Xiao Zhan}, \bibinfo{person}{Yaxiong Lei}, \bibinfo{person}{Yueyan Liu}, \bibinfo{person}{Ruba Abu-Salma}, {and} \bibinfo{person}{Jose Such}.} \bibinfo{year}{2025}\natexlab{}.
\newblock \showarticletitle{Exploring the privacy and security challenges faced by migrant domestic workers in chinese smart homes}. In \bibinfo{booktitle}{\emph{Proceedings of the 2025 CHI Conference on Human Factors in Computing Systems}}. \bibinfo{pages}{1--18}.
\newblock


\bibitem[Hu et~al\mbox{.}(2024)]%
        {hu2024exploring}
\bibfield{author}{\bibinfo{person}{Yongquan Hu}, \bibinfo{person}{Shuning Zhang}, \bibinfo{person}{Ting Dang}, \bibinfo{person}{Hong Jia}, \bibinfo{person}{Flora~D Salim}, \bibinfo{person}{Wen Hu}, {and} \bibinfo{person}{Aaron~J Quigley}.} \bibinfo{year}{2024}\natexlab{}.
\newblock \showarticletitle{Exploring large-scale language models to evaluate eeg-based multimodal data for mental health}. In \bibinfo{booktitle}{\emph{Companion of the 2024 on ACM International Joint Conference on Pervasive and Ubiquitous Computing}}. \bibinfo{pages}{412--417}.
\newblock


\bibitem[iOS(2022)]%
        {apple_ios_2022}
\bibfield{author}{\bibinfo{person}{Apple iOS}.} \bibinfo{year}{2022}\natexlab{}.
\newblock \bibinfo{title}{Accessing private data}.
\newblock
\urldef\tempurl%
\url{https://developer.apple.com/design/human-interface-guidelines/patterns/accessing-private-data/}
\showURL{%
\tempurl}
\newblock
\shownote{Accessed: 2025-01-14}.


\bibitem[Jana et~al\mbox{.}(2013)]%
        {jana2013enabling}
\bibfield{author}{\bibinfo{person}{Suman Jana}, \bibinfo{person}{David Molnar}, \bibinfo{person}{Alexander Moshchuk}, \bibinfo{person}{Alan Dunn}, \bibinfo{person}{Benjamin Livshits}, \bibinfo{person}{Helen~J Wang}, {and} \bibinfo{person}{Eyal Ofek}.} \bibinfo{year}{2013}\natexlab{}.
\newblock \showarticletitle{Enabling $\{$Fine-Grained$\}$ permissions for augmented reality applications with recognizers}. In \bibinfo{booktitle}{\emph{22nd USENIX Security Symposium (USENIX Security 13)}}. \bibinfo{pages}{415--430}.
\newblock


\bibitem[Kumaraguru and Cranor(2005)]%
        {kumaraguru2005privacy}
\bibfield{author}{\bibinfo{person}{Ponnurangam Kumaraguru} {and} \bibinfo{person}{Lorrie~Faith Cranor}.} \bibinfo{year}{2005}\natexlab{}.
\newblock \showarticletitle{Privacy indexes: a survey of Westin's studies}.
\newblock  (\bibinfo{year}{2005}).
\newblock


\bibitem[Kuznetsova et~al\mbox{.}(2020)]%
        {kuznetsova2020open}
\bibfield{author}{\bibinfo{person}{Alina Kuznetsova}, \bibinfo{person}{Hassan Rom}, \bibinfo{person}{Neil Alldrin}, \bibinfo{person}{Jasper Uijlings}, \bibinfo{person}{Ivan Krasin}, \bibinfo{person}{Jordi Pont-Tuset}, \bibinfo{person}{Shahab Kamali}, \bibinfo{person}{Stefan Popov}, \bibinfo{person}{Matteo Malloci}, \bibinfo{person}{Alexander Kolesnikov}, {et~al\mbox{.}}} \bibinfo{year}{2020}\natexlab{}.
\newblock \showarticletitle{The open images dataset v4: Unified image classification, object detection, and visual relationship detection at scale}.
\newblock \bibinfo{journal}{\emph{International journal of computer vision}} \bibinfo{volume}{128}, \bibinfo{number}{7} (\bibinfo{year}{2020}), \bibinfo{pages}{1956--1981}.
\newblock


\bibitem[Lee et~al\mbox{.}(2024)]%
        {lee2024priviaware}
\bibfield{author}{\bibinfo{person}{Hyunsoo Lee}, \bibinfo{person}{Yugyeong Jung}, \bibinfo{person}{Hei~Yiu Law}, \bibinfo{person}{Seolyeong Bae}, {and} \bibinfo{person}{Uichin Lee}.} \bibinfo{year}{2024}\natexlab{}.
\newblock \showarticletitle{PriviAware: Exploring Data Visualization and Dynamic Privacy Control Support for Data Collection in Mobile Sensing Research}. In \bibinfo{booktitle}{\emph{Proceedings of the CHI Conference on Human Factors in Computing Systems}}. \bibinfo{pages}{1--17}.
\newblock


\bibitem[Li et~al\mbox{.}(2023)]%
        {li2023occupant}
\bibfield{author}{\bibinfo{person}{Beatrice Li}, \bibinfo{person}{Arash Tavakoli}, {and} \bibinfo{person}{Arsalan Heydarian}.} \bibinfo{year}{2023}\natexlab{}.
\newblock \showarticletitle{Occupant privacy perception, awareness, and preferences in smart office environments}.
\newblock \bibinfo{journal}{\emph{Scientific Reports}} \bibinfo{volume}{13}, \bibinfo{number}{1} (\bibinfo{year}{2023}), \bibinfo{pages}{4073}.
\newblock


\bibitem[Liu et~al\mbox{.}(2022)]%
        {liu2022development}
\bibfield{author}{\bibinfo{person}{Donghang Liu}, \bibinfo{person}{Chensi Wu}, \bibinfo{person}{Lulin Yang}, \bibinfo{person}{Xiaoying Zhao}, {and} \bibinfo{person}{Qifeng Sun}.} \bibinfo{year}{2022}\natexlab{}.
\newblock \showarticletitle{The development of privacy protection standards for smart home}.
\newblock \bibinfo{journal}{\emph{Wireless Communications and Mobile Computing}} \bibinfo{volume}{2022}, \bibinfo{number}{1} (\bibinfo{year}{2022}), \bibinfo{pages}{9641143}.
\newblock


\bibitem[Ma et~al\mbox{.}(2024)]%
        {ma2024understanding}
\bibfield{author}{\bibinfo{person}{Ying Ma}, \bibinfo{person}{Cherie Sew}, \bibinfo{person}{Zhanna Sarsenbayeva}, \bibinfo{person}{Jarrod Knibbe}, {and} \bibinfo{person}{Jorge Goncalves}.} \bibinfo{year}{2024}\natexlab{}.
\newblock \showarticletitle{Understanding Users' Perspectives on Location Privacy Management on iPhones}.
\newblock \bibinfo{journal}{\emph{Proceedings of the ACM on Human-Computer Interaction}} \bibinfo{volume}{8}, \bibinfo{number}{MHCI} (\bibinfo{year}{2024}), \bibinfo{pages}{1--25}.
\newblock


\bibitem[Ma et~al\mbox{.}(2025)]%
        {ma2025raising}
\bibfield{author}{\bibinfo{person}{Ying Ma}, \bibinfo{person}{Shiquan Zhang}, \bibinfo{person}{Dongju Yang}, \bibinfo{person}{Zhanna Sarsenbayeva}, \bibinfo{person}{Jarrod Knibbe}, {and} \bibinfo{person}{Jorge Goncalves}.} \bibinfo{year}{2025}\natexlab{}.
\newblock \showarticletitle{Raising Awareness of Location Information Vulnerabilities in Social Media Photos using LLMs}. In \bibinfo{booktitle}{\emph{Proceedings of the 2025 CHI Conference on Human Factors in Computing Systems}}. \bibinfo{pages}{1--14}.
\newblock


\bibitem[Mahmud et~al\mbox{.}(2024)]%
        {mahmud2024actsonic}
\bibfield{author}{\bibinfo{person}{Saif Mahmud}, \bibinfo{person}{Vineet Parikh}, \bibinfo{person}{Qikang Liang}, \bibinfo{person}{Ke Li}, \bibinfo{person}{Ruidong Zhang}, \bibinfo{person}{Ashwin Ajit}, \bibinfo{person}{Vipin Gunda}, \bibinfo{person}{Devansh Agarwal}, \bibinfo{person}{Fran{\c{c}}ois Guimbreti{\`e}re}, {and} \bibinfo{person}{Cheng Zhang}.} \bibinfo{year}{2024}\natexlab{}.
\newblock \showarticletitle{ActSonic: Recognizing Everyday Activities from Inaudible Acoustic Wave Around the Body}.
\newblock \bibinfo{journal}{\emph{Proceedings of the ACM on Interactive, Mobile, Wearable and Ubiquitous Technologies}} \bibinfo{volume}{8}, \bibinfo{number}{4} (\bibinfo{year}{2024}), \bibinfo{pages}{1--32}.
\newblock


\bibitem[Nair et~al\mbox{.}(2023)]%
        {nair2023going}
\bibfield{author}{\bibinfo{person}{Vivek~C Nair}, \bibinfo{person}{Gonzalo Munilla-Garrido}, {and} \bibinfo{person}{Dawn Song}.} \bibinfo{year}{2023}\natexlab{}.
\newblock \showarticletitle{Going incognito in the metaverse: Achieving theoretically optimal privacy-usability tradeoffs in VR}. In \bibinfo{booktitle}{\emph{Proceedings of the 36th Annual ACM Symposium on User Interface Software and Technology}}. \bibinfo{pages}{1--16}.
\newblock


\bibitem[O'Hagan et~al\mbox{.}(2023)]%
        {o2023privacy}
\bibfield{author}{\bibinfo{person}{Joseph O'Hagan}, \bibinfo{person}{Pejman Saeghe}, \bibinfo{person}{Jan Gugenheimer}, \bibinfo{person}{Daniel Medeiros}, \bibinfo{person}{Karola Marky}, \bibinfo{person}{Mohamed Khamis}, {and} \bibinfo{person}{Mark McGill}.} \bibinfo{year}{2023}\natexlab{}.
\newblock \showarticletitle{Privacy-enhancing technology and everyday augmented reality: Understanding bystanders' varying needs for awareness and consent}.
\newblock \bibinfo{journal}{\emph{Proceedings of the ACM on Interactive, Mobile, Wearable and Ubiquitous Technologies}} \bibinfo{volume}{6}, \bibinfo{number}{4} (\bibinfo{year}{2023}), \bibinfo{pages}{1--35}.
\newblock


\bibitem[Orekondy et~al\mbox{.}(2017)]%
        {orekondy2017towards}
\bibfield{author}{\bibinfo{person}{Tribhuvanesh Orekondy}, \bibinfo{person}{Bernt Schiele}, {and} \bibinfo{person}{Mario Fritz}.} \bibinfo{year}{2017}\natexlab{}.
\newblock \showarticletitle{Towards a visual privacy advisor: Understanding and predicting privacy risks in images}. In \bibinfo{booktitle}{\emph{Proceedings of the IEEE international conference on computer vision}}. \bibinfo{pages}{3686--3695}.
\newblock


\bibitem[Prange et~al\mbox{.}(2021)]%
        {prange2021priview}
\bibfield{author}{\bibinfo{person}{Sarah Prange}, \bibinfo{person}{Ahmed Shams}, \bibinfo{person}{Robin Piening}, \bibinfo{person}{Yomna Abdelrahman}, {and} \bibinfo{person}{Florian Alt}.} \bibinfo{year}{2021}\natexlab{}.
\newblock \showarticletitle{Priview--exploring visualisations to support users’ privacy awareness}. In \bibinfo{booktitle}{\emph{Proceedings of the 2021 chi conference on human factors in computing systems}}. \bibinfo{pages}{1--18}.
\newblock


\bibitem[Raval et~al\mbox{.}(2016)]%
        {raval2016you}
\bibfield{author}{\bibinfo{person}{Nisarg Raval}, \bibinfo{person}{Animesh Srivastava}, \bibinfo{person}{Ali Razeen}, \bibinfo{person}{Kiron Lebeck}, \bibinfo{person}{Ashwin Machanavajjhala}, {and} \bibinfo{person}{Lanodn~P Cox}.} \bibinfo{year}{2016}\natexlab{}.
\newblock \showarticletitle{What you mark is what apps see}. In \bibinfo{booktitle}{\emph{Proceedings of the 14th Annual International Conference on Mobile Systems, Applications, and Services}}. \bibinfo{pages}{249--261}.
\newblock


\bibitem[Shu et~al\mbox{.}(2018)]%
        {shu2018cardea}
\bibfield{author}{\bibinfo{person}{Jiayu Shu}, \bibinfo{person}{Rui Zheng}, {and} \bibinfo{person}{Pan Hui}.} \bibinfo{year}{2018}\natexlab{}.
\newblock \showarticletitle{Cardea: Context-aware visual privacy protection for photo taking and sharing}. In \bibinfo{booktitle}{\emph{Proceedings of the 9th ACM Multimedia Systems Conference}}. \bibinfo{pages}{304--315}.
\newblock


\bibitem[Song et~al\mbox{.}(2020)]%
        {song2020m}
\bibfield{author}{\bibinfo{person}{Yunpeng Song}, \bibinfo{person}{Yun Huang}, \bibinfo{person}{Zhongmin Cai}, {and} \bibinfo{person}{Jason~I Hong}.} \bibinfo{year}{2020}\natexlab{}.
\newblock \showarticletitle{I'm all eyes and ears: Exploring effective locators for privacy awareness in iot scenarios}. In \bibinfo{booktitle}{\emph{Proceedings of the 2020 CHI Conference on Human Factors in Computing Systems}}. \bibinfo{pages}{1--13}.
\newblock


\bibitem[Steil et~al\mbox{.}(2019)]%
        {steil2019privaceye}
\bibfield{author}{\bibinfo{person}{Julian Steil}, \bibinfo{person}{Marion Koelle}, \bibinfo{person}{Wilko Heuten}, \bibinfo{person}{Susanne Boll}, {and} \bibinfo{person}{Andreas Bulling}.} \bibinfo{year}{2019}\natexlab{}.
\newblock \showarticletitle{Privaceye: privacy-preserving head-mounted eye tracking using egocentric scene image and eye movement features}. In \bibinfo{booktitle}{\emph{Proceedings of the 11th ACM symposium on eye tracking research \& applications}}. \bibinfo{pages}{1--10}.
\newblock


\bibitem[T{\"o}mek{\c{c}}e et~al\mbox{.}({[n.\,d.]})]%
        {tomekcceprivate}
\bibfield{author}{\bibinfo{person}{Batuhan T{\"o}mek{\c{c}}e}, \bibinfo{person}{Mark Vero}, \bibinfo{person}{Robin Staab}, {and} \bibinfo{person}{Martin Vechev}.} \bibinfo{year}{[n.\,d.]}\natexlab{}.
\newblock \showarticletitle{Private Attribute Inference from Images with Vision-Language Models}. In \bibinfo{booktitle}{\emph{The Thirty-eighth Annual Conference on Neural Information Processing Systems}}.
\newblock


\bibitem[Venugopalan et~al\mbox{.}(2024)]%
        {venugopalan2024aragorn}
\bibfield{author}{\bibinfo{person}{Hari Venugopalan}, \bibinfo{person}{Zainul~Abi Din}, \bibinfo{person}{Trevor Carpenter}, \bibinfo{person}{Jason Lowe-Power}, \bibinfo{person}{Samuel~T King}, {and} \bibinfo{person}{Zubair Shafiq}.} \bibinfo{year}{2024}\natexlab{}.
\newblock \showarticletitle{Aragorn: A Privacy-Enhancing System for Mobile Cameras}.
\newblock \bibinfo{journal}{\emph{Proceedings of the ACM on Interactive, Mobile, Wearable and Ubiquitous Technologies}} \bibinfo{volume}{7}, \bibinfo{number}{4} (\bibinfo{year}{2024}), \bibinfo{pages}{1--31}.
\newblock


\bibitem[Wang et~al\mbox{.}(2023)]%
        {wang2023modeling}
\bibfield{author}{\bibinfo{person}{Yuntao Wang}, \bibinfo{person}{Zirui Cheng}, \bibinfo{person}{Xin Yi}, \bibinfo{person}{Yan Kong}, \bibinfo{person}{Xueyang Wang}, \bibinfo{person}{Xuhai Xu}, \bibinfo{person}{Yukang Yan}, \bibinfo{person}{Chun Yu}, \bibinfo{person}{Shwetak Patel}, {and} \bibinfo{person}{Yuanchun Shi}.} \bibinfo{year}{2023}\natexlab{}.
\newblock \showarticletitle{Modeling the trade-off of privacy preservation and activity recognition on low-resolution images}. In \bibinfo{booktitle}{\emph{Proceedings of the 2023 CHI Conference on Human Factors in Computing Systems}}. \bibinfo{pages}{1--15}.
\newblock


\bibitem[Wijesekera et~al\mbox{.}(2017)]%
        {wijesekera2017feasibility}
\bibfield{author}{\bibinfo{person}{Primal Wijesekera}, \bibinfo{person}{Arjun Baokar}, \bibinfo{person}{Lynn Tsai}, \bibinfo{person}{Joel Reardon}, \bibinfo{person}{Serge Egelman}, \bibinfo{person}{David Wagner}, {and} \bibinfo{person}{Konstantin Beznosov}.} \bibinfo{year}{2017}\natexlab{}.
\newblock \showarticletitle{The feasibility of dynamically granted permissions: Aligning mobile privacy with user preferences}. In \bibinfo{booktitle}{\emph{2017 IEEE Symposium on Security and Privacy (SP)}}. IEEE, \bibinfo{pages}{1077--1093}.
\newblock


\bibitem[Wijesekera et~al\mbox{.}(2018)]%
        {wijesekera2018contextualizing}
\bibfield{author}{\bibinfo{person}{Primal Wijesekera}, \bibinfo{person}{Joel Reardon}, \bibinfo{person}{Irwin Reyes}, \bibinfo{person}{Lynn Tsai}, \bibinfo{person}{Jung-Wei Chen}, \bibinfo{person}{Nathan Good}, \bibinfo{person}{David Wagner}, \bibinfo{person}{Konstantin Beznosov}, {and} \bibinfo{person}{Serge Egelman}.} \bibinfo{year}{2018}\natexlab{}.
\newblock \showarticletitle{Contextualizing privacy decisions for better prediction (and protection)}. In \bibinfo{booktitle}{\emph{Proceedings of the 2018 CHI Conference on Human Factors in Computing Systems}}. \bibinfo{pages}{1--13}.
\newblock


\bibitem[Wu et~al\mbox{.}(2020)]%
        {wu2020privacy}
\bibfield{author}{\bibinfo{person}{Zhenyu Wu}, \bibinfo{person}{Haotao Wang}, \bibinfo{person}{Zhaowen Wang}, \bibinfo{person}{Hailin Jin}, {and} \bibinfo{person}{Zhangyang Wang}.} \bibinfo{year}{2020}\natexlab{}.
\newblock \showarticletitle{Privacy-preserving deep action recognition: An adversarial learning framework and a new dataset}.
\newblock \bibinfo{journal}{\emph{IEEE Transactions on Pattern Analysis and Machine Intelligence}} \bibinfo{volume}{44}, \bibinfo{number}{4} (\bibinfo{year}{2020}), \bibinfo{pages}{2126--2139}.
\newblock


\bibitem[Xu et~al\mbox{.}(2023)]%
        {xu2023dipa}
\bibfield{author}{\bibinfo{person}{Anran Xu}, \bibinfo{person}{Zhongyi Zhou}, \bibinfo{person}{Kakeru Miyazaki}, \bibinfo{person}{Ryo Yoshikawa}, \bibinfo{person}{Simo Hosio}, {and} \bibinfo{person}{Koji Yatani}.} \bibinfo{year}{2023}\natexlab{}.
\newblock \showarticletitle{DIPA: An Image Dataset with Cross-cultural Privacy Concern Annotations}. In \bibinfo{booktitle}{\emph{Companion Proceedings of the 28th International Conference on Intelligent User Interfaces}}. \bibinfo{pages}{259--266}.
\newblock


\bibitem[Xu et~al\mbox{.}(2024)]%
        {xu2024dipa2}
\bibfield{author}{\bibinfo{person}{Anran Xu}, \bibinfo{person}{Zhongyi Zhou}, \bibinfo{person}{Kakeru Miyazaki}, \bibinfo{person}{Ryo Yoshikawa}, \bibinfo{person}{Simo Hosio}, {and} \bibinfo{person}{Koji Yatani}.} \bibinfo{year}{2024}\natexlab{}.
\newblock \showarticletitle{DIPA2: An Image Dataset with Cross-cultural Privacy Perception Annotations}.
\newblock \bibinfo{journal}{\emph{Proceedings of the ACM on Interactive, Mobile, Wearable and Ubiquitous Technologies}} \bibinfo{volume}{7}, \bibinfo{number}{4} (\bibinfo{year}{2024}), \bibinfo{pages}{1--30}.
\newblock


\bibitem[Zhang and Sundar(2019)]%
        {zhang2019proactive}
\bibfield{author}{\bibinfo{person}{Bo Zhang} {and} \bibinfo{person}{S~Shyam Sundar}.} \bibinfo{year}{2019}\natexlab{}.
\newblock \showarticletitle{Proactive vs. reactive personalization: Can customization of privacy enhance user experience?}
\newblock \bibinfo{journal}{\emph{International journal of human-computer studies}}  \bibinfo{volume}{128} (\bibinfo{year}{2019}), \bibinfo{pages}{86--99}.
\newblock


\bibitem[Zhang et~al\mbox{.}(2025b)]%
        {zhang2025interveeg}
\bibfield{author}{\bibinfo{person}{Shuning Zhang}, \bibinfo{person}{Yongquan'Owen' Hu}, \bibinfo{person}{Xin Yi}, \bibinfo{person}{Suranga Nanayakkara}, {and} \bibinfo{person}{Xiaoming Chen}.} \bibinfo{year}{2025}\natexlab{b}.
\newblock \showarticletitle{IntervEEG-LLM: Exploring EEG-Based Multimodal Data for Customized Mental Health Interventions}. In \bibinfo{booktitle}{\emph{Companion Proceedings of the ACM on Web Conference 2025}}. \bibinfo{pages}{2320--2326}.
\newblock


\bibitem[Zhang et~al\mbox{.}(2024b)]%
        {zhang2024ghost}
\bibfield{author}{\bibinfo{person}{Shuning Zhang}, \bibinfo{person}{Lyumanshan Ye}, \bibinfo{person}{Xin Yi}, \bibinfo{person}{Jingyu Tang}, \bibinfo{person}{Bo Shui}, \bibinfo{person}{Haobin Xing}, \bibinfo{person}{Pengfei Liu}, {and} \bibinfo{person}{Hewu Li}.} \bibinfo{year}{2024}\natexlab{b}.
\newblock \showarticletitle{" Ghost of the past": identifying and resolving privacy leakage from LLM's memory through proactive user interaction}.
\newblock \bibinfo{journal}{\emph{arXiv preprint arXiv:2410.14931}} (\bibinfo{year}{2024}).
\newblock


\bibitem[Zhang et~al\mbox{.}(2025c)]%
        {zhang2025privcaptcha}
\bibfield{author}{\bibinfo{person}{Shuning Zhang}, \bibinfo{person}{Xin Yi}, \bibinfo{person}{Shixuan Li}, \bibinfo{person}{Haobin Xing}, {and} \bibinfo{person}{Hewu Li}.} \bibinfo{year}{2025}\natexlab{c}.
\newblock \showarticletitle{PrivCAPTCHA: Interactive CAPTCHA to Facilitate Effective Comprehension of APP Privacy Policy}. In \bibinfo{booktitle}{\emph{Proceedings of the 2025 CHI Conference on Human Factors in Computing Systems}}. \bibinfo{pages}{1--20}.
\newblock


\bibitem[Zhang et~al\mbox{.}(2024c)]%
        {zhang2024adanonymizer}
\bibfield{author}{\bibinfo{person}{Shuning Zhang}, \bibinfo{person}{Xin Yi}, \bibinfo{person}{Haobin Xing}, \bibinfo{person}{Lyumanshan Ye}, \bibinfo{person}{Yongquan Hu}, {and} \bibinfo{person}{Hewu Li}.} \bibinfo{year}{2024}\natexlab{c}.
\newblock \showarticletitle{Adanonymizer: Interactively Navigating and Balancing the Duality of Privacy and Output Performance in Human-LLM Interaction}.
\newblock \bibinfo{journal}{\emph{arXiv preprint arXiv:2410.15044}} (\bibinfo{year}{2024}).
\newblock


\bibitem[Zhang et~al\mbox{.}(2025a)]%
        {zhang2025through}
\bibfield{author}{\bibinfo{person}{Ziyang Zhang}, \bibinfo{person}{Chong Bao}, \bibinfo{person}{Xiaokun Pan}, \bibinfo{person}{Chia-Ming Chang}, \bibinfo{person}{Takeo Igarashi}, {and} \bibinfo{person}{Guofeng Zhang}.} \bibinfo{year}{2025}\natexlab{a}.
\newblock \showarticletitle{Through the Lens of Privacy: Exploring Privacy Protection in Vision-Language Model Interactions on Smart Glasses}. In \bibinfo{booktitle}{\emph{Proceedings of the Extended Abstracts of the CHI Conference on Human Factors in Computing Systems}}. \bibinfo{pages}{1--8}.
\newblock


\bibitem[Zhang et~al\mbox{.}(2024a)]%
        {zhang2024privacy}
\bibfield{author}{\bibinfo{person}{Zhiping Zhang}, \bibinfo{person}{Bingcan Guo}, {and} \bibinfo{person}{Tianshi Li}.} \bibinfo{year}{2024}\natexlab{a}.
\newblock \showarticletitle{Privacy Leakage Overshadowed by Views of AI: A Study on Human Oversight of Privacy in Language Model Agent}.
\newblock \bibinfo{journal}{\emph{arXiv preprint arXiv:2411.01344}} (\bibinfo{year}{2024}).
\newblock


\bibitem[Zhao et~al\mbox{.}(2025)]%
        {zhao2025visual}
\bibfield{author}{\bibinfo{person}{Ruoyu Zhao}, \bibinfo{person}{Yushu Zhang}, \bibinfo{person}{Tao Wang}, \bibinfo{person}{Wenying Wen}, \bibinfo{person}{Yong Xiang}, {and} \bibinfo{person}{Xiaochun Cao}.} \bibinfo{year}{2025}\natexlab{}.
\newblock \showarticletitle{Visual content privacy protection: A survey}.
\newblock \bibinfo{journal}{\emph{Comput. Surveys}} \bibinfo{volume}{57}, \bibinfo{number}{5} (\bibinfo{year}{2025}), \bibinfo{pages}{1--36}.
\newblock


\bibitem[Zheng et~al\mbox{.}(2023)]%
        {zheng2023judging}
\bibfield{author}{\bibinfo{person}{Lianmin Zheng}, \bibinfo{person}{Wei-Lin Chiang}, \bibinfo{person}{Ying Sheng}, \bibinfo{person}{Siyuan Zhuang}, \bibinfo{person}{Zhanghao Wu}, \bibinfo{person}{Yonghao Zhuang}, \bibinfo{person}{Zi Lin}, \bibinfo{person}{Zhuohan Li}, \bibinfo{person}{Dacheng Li}, \bibinfo{person}{Eric Xing}, {et~al\mbox{.}}} \bibinfo{year}{2023}\natexlab{}.
\newblock \showarticletitle{Judging llm-as-a-judge with mt-bench and chatbot arena}.
\newblock \bibinfo{journal}{\emph{Advances in Neural Information Processing Systems}}  \bibinfo{volume}{36} (\bibinfo{year}{2023}), \bibinfo{pages}{46595--46623}.
\newblock


\bibitem[Zheng et~al\mbox{.}(2018)]%
        {zheng2018user}
\bibfield{author}{\bibinfo{person}{Serena Zheng}, \bibinfo{person}{Noah Apthorpe}, \bibinfo{person}{Marshini Chetty}, {and} \bibinfo{person}{Nick Feamster}.} \bibinfo{year}{2018}\natexlab{}.
\newblock \showarticletitle{User perceptions of smart home IoT privacy}.
\newblock \bibinfo{journal}{\emph{Proceedings of the ACM on human-computer interaction}} \bibinfo{volume}{2}, \bibinfo{number}{CSCW} (\bibinfo{year}{2018}), \bibinfo{pages}{1--20}.
\newblock


\bibitem[Zhou et~al\mbox{.}(2024)]%
        {zhou2024bring}
\bibfield{author}{\bibinfo{person}{Haozhe Zhou}, \bibinfo{person}{Mayank Goel}, {and} \bibinfo{person}{Yuvraj Agarwal}.} \bibinfo{year}{2024}\natexlab{}.
\newblock \showarticletitle{Bring Privacy To The Table: Interactive Negotiation for Privacy Settings of Shared Sensing Devices}. In \bibinfo{booktitle}{\emph{Proceedings of the 2024 CHI Conference on Human Factors in Computing Systems}}. \bibinfo{pages}{1--22}.
\newblock


\bibitem[Zhou et~al\mbox{.}(2016)]%
        {zhou2016enhancing}
\bibfield{author}{\bibinfo{person}{Huiyuan Zhou}, \bibinfo{person}{Khalid Tearo}, \bibinfo{person}{Aniruddha Waje}, \bibinfo{person}{Elham Alghamdi}, \bibinfo{person}{Thamara Alves}, \bibinfo{person}{Vinicius Ferreira}, \bibinfo{person}{Kirstie Hawkey}, {and} \bibinfo{person}{Derek Reilly}.} \bibinfo{year}{2016}\natexlab{}.
\newblock \showarticletitle{Enhancing mobile content privacy with proxemics aware notifications and protection}. In \bibinfo{booktitle}{\emph{Proceedings of the 2016 CHI Conference on Human Factors in Computing Systems}}. \bibinfo{pages}{1362--1373}.
\newblock


\end{thebibliography}

\appendix

\section{Prompt Structure}

\end{document}